\documentclass[letterpaper, 10 pt, conference]{ieeeconf}  

\IEEEoverridecommandlockouts                              
\usepackage{cite}
\usepackage{amsmath,amssymb,amsfonts}
\usepackage{algorithmic}
\usepackage{graphicx}
\usepackage{textcomp}

\usepackage{subfigure}
\usepackage{epsfig} 
\usepackage{hyperref}
\usepackage{booktabs}
\usepackage{bm}
\usepackage{bbm}
\usepackage{mathtools}
\usepackage{xcolor}

\title{\LARGE \bf
Amortized Nonlinear Model Predictive Control
}

\newcommand{\R}{\mathbb{R}}

\newcommand{\x}{x}
\newcommand{\uv}{u}
\newcommand{\wv}{w}
\newcommand{\zv}{z}
\newcommand{\sv}{s}
\newcommand{\lam}{\lambda}
\newcommand{\mum}{\mu}
\newcommand{\xref}{x_{\mathrm{ref}}}
\newcommand{\uref}{u_{\mathrm{ref}}}
\newcommand{\yref}{y_{\mathrm{ref}}}
\newcommand{\calX}{\mathcal{X}}
\newcommand{\calU}{\mathcal{U}}
\newcommand{\calD}{\mathcal{D}}
\newcommand{\calL}{\mathcal{L}}

\newcommand{\norm}[1]{\left\lVert #1 \right\rVert}
\newcommand{\transpose}{{{}^{\top}}}
\newcommand{\umin}{\underline{u}}
\newcommand{\umax}{\overline{u}}
\newcommand{\xmin}{\underline{x}}
\newcommand{\xmax}{\overline{x}}

\newtheorem{theorem}{Theorem}[section]

\newtheorem{remark}[theorem]{Remark}

\author{Francesco Pillitteri,
Alberto Bemporad
\thanks{This work was funded by the European Union (ERC Advanced Research Grant COMPACT, No. 101141351). Views and opinions expressed are however those of the authors only and do not necessarily reflect those of the European Union or the European Research Council. Neither the European Union nor the granting authority can be held responsible for them.}
\thanks{The authors are with the IMT School for Advanced Studies, Lucca, Italy. Email: \texttt{francesco.pillitteri@imtlucca.it}}
}

\begin{document}


\maketitle
\thispagestyle{empty}
\pagestyle{empty}

\begin{abstract}
  Nonlinear Model Predictive Control requires solving a constrained nonlinear program (NLP) in real-time at every sampling instant,
  a computational bottleneck that limits deployment on resource-constrained
  hardware or at high sampling rates. We address this challenge for the broad class
  of \emph{input-affine} nonlinear systems to show that the optimal control move can be
  approximated by a \emph{state-dependent quadratic program} (QP) whose
  cost parameters depend on the current state and reference.
  We propose a single-network \emph{residual-corrector} architecture: a
  state-dependent analytic baseline provides initial QP parameters, and the
  network learns only the corrections needed to match the full NLP solution; the
  QP is solved by a differentiable interior-point layer, guaranteeing
  constraint satisfaction for the first control action.  The network is
  trained offline on data generated by an NLP solver using a
  hybrid loss that combines supervised imitation and KKT-residual penalties.
  We validate the approach on a three-link planar
  robotic arm with Cartesian end-effector tracking, demonstrating
  orders-of-magnitude speedup over the NLP solver while maintaining
  comparable tracking performance.
\end{abstract}


\section{Introduction}
\label{sec:intro}

Model Predictive Control (MPC) is widely recognized as a powerful approach to
constrained optimal control: at each time step, the controller solves a
finite-horizon optimization problem, applies the first element of the
optimal input sequence, and repeats the process at the next
step~\cite{rawlings2017model}. From the perspective of optimization
theory, MPC is inherently a \emph{parametric} program: the current
state $\x_k$ (and any output reference vector) acts as the parameter, and the
optimal first input $\uv_0^\star(\x_k)$ is the corresponding solution
map. For linear systems with
quadratic costs and polyhedral constraints the resulting program is a
\emph{convex QP} that can be solved in milliseconds and less 
by modern active-set~\cite{ferreau2014qpoases,ODYSQP}
or interior-point solvers~\cite{frison2020hpipm}. An explicit
piecewise-affine solution map can be even pre-computed offline via
\emph{multi-parametric programming}~\cite{bemporad2002explicit}
to avoid online optimization altogether.

For nonlinear plant models, the problem is a nonlinear program (NLP), whose
per-step solve time may be orders of magnitude larger, often rendering
real-time implementation infeasible, and explicit nonlinear MPC is in general 
intractable. A classical workaround is to perform only a single Newton-type iteration
per step, warm-starting from the previous solution shifted by one time step. This underpins
\emph{real-time iteration} (RTI) schemes~\cite{diehl2002real,gros2020linear},
which have proved effective in many applications, though the per-step cost
can still be prohibitive at high sampling rates or on resource-constrained
hardware.

Recent advances in machine learning offer an alternative route: use 
function approximation to replace the online solver, in analogy to explicit MPC, 
or to substantially accelerate it by providing a good initial guess. This idea, also referred to as \emph{amortized optimization}~\cite{amos2022tutorial}, has attracted growing interest for MPC due to the microsecond inference times it
enables once the approximation is trained offline. A related approach to reduce the online computational burden in nonlinear MPC is to approximate the cost-to-go function, see, e.g.,~\cite{AZB24,SBB25}.

Due to the approximation, preserving \emph{closed-loop guarantees} can be a challenge, and several strategies have been explored in the literature.
For example, stability and feasibility can be recovered by augmenting the approximator
with a safety filter~\cite{wabersich2021predictive}.
Alternatively, learning a warm-start rather than the full solution
preserves problem feasibility while reducing iteration
counts~\cite{sambharya2023end}. Differentiable optimization layers embed
a QP as a differentiable mapping inside the network, guaranteeing that
the output satisfies the QP constraints by
construction~\cite{amos2017optnet}. Our method belongs to this last family.

\textbf{Contribution.} 
Rather than learning the full optimal
control sequence, we learn a network that predicts the parameters of a
\emph{small QP} optimizing the first control move, which is then solved exactly by a
differentiable layer. This separates the learning problem (predicting
good QP-problem matrices) from constraint enforcement (handled by the solver). 
After formalizing the MPC problem for general input-affine nonlinear systems and showing how the input-affine structure yields a \emph{parametric QP} for the first control step,
we propose a \emph{residual-corrector QP architecture}, which learns how to correct state-dependent QP baseline obtained analytically. Hard constraints are enforced by the QP solver, while state constraints are softened to avoid infeasibility. The network is trained offline on NLP solutions via a \emph{hybrid KKT loss} combining supervised imitation and KKT-residual penalties. We validate the approach on a three-link planar robotic arm with Cartesian end-effector tracking, demonstrating orders-of-magnitude speedup over the NLP solver while maintaining comparable tracking performance and constraint satisfaction.

The paper is organized as follows. Section~\ref{sec:problem} formulates the MPC problem for input-affine systems. Section~\ref{sec:method} presents the proposed architecture and training
procedure.
Section~\ref{sec:experiments} reports experimental results on the robotic arm.
Section~\ref{sec:conclusion} concludes the paper.

\emph{Notation.}
We denote by $I_n$ the identity matrix of order $n$ and by $\mathrm{diag}(v)$ the diagonal
matrix formed from vector $v$.  For a matrix $Q=Q'\succeq 0$,
$\|a\|_Q^2 := a\transpose Q\,a$ and for $Q\succ 0$ we denote by $L=\mathrm{chol}(Q)$ its lower-triangular Cholesky factor, $Q = L L\transpose$. The operator $[\cdot]^+\!=\!\max(\cdot,0)$ and
$[\cdot]^-\!=\!\max(-\cdot,0)$ extract the positive and negative parts,
respectively. 

\section{Problem Formulation}
\label{sec:problem}

\subsection{Input-Affine Nonlinear Models}

We consider discrete-time input-affine nonlinear prediction models of the form
\begin{equation}
  \x_{k+1} = f(\x_k) + G(\x_k)\,\uv_k,
  \label{eq:dynamics}
\end{equation}
where $\x_k\in \R^{n_x}$ is the state vector, 
$\uv_k \in \R^{n_u}$ the control input, and $f : \R^{n_x} \to \R^{n_x}$ and 
$G : \R^{n_x} \to \R^{n_x \times n_u}$ define the state-update equation. 
Affine systems~\eqref{eq:dynamics} either arise naturally in many engineering domains, such as in robotics; at the price of introducing an input delay, any general dynamics $\x_{k+1}=F(\x_{k},\uv_k)$ can be always transformed into affine form by treating $\uv_{k}$ as an additional state.

We assume that~\eqref{eq:dynamics} is subject to constraints, namely \emph{box constraints}
\begin{equation}
  \begin{aligned}
  & \calU = \{\uv \in \R^{n_u} : \umin \leq \uv \leq \umax\}, \\
  & \calX = \{\x \in \R^{n_x} : \xmin \leq \x \leq \xmax\},
  \end{aligned}
  \label{eq:box_constraints}
\end{equation}
and \emph{affine-in-input} constraints
\begin{equation}
  A_c(\x_k)\,\uv_k \leq b_c(\x_k),
  \label{eq:affine_input_constr}
\end{equation}
where $A_c(\x_k)\in\R^{n_c\times n_u}$ and $b_c(\x_k)\in\R^{n_c}$.  
Constraints of the form~\eqref{eq:affine_input_constr} capture, for example, joint-torque
limits expressed via inverse dynamics
($\tau = M(q)\uv + h(q,\dot{q})$), as well as
mixed state/input polytopic constraints that become affine in $\uv_k$ for any given
$\x_k$.

\subsection{Receding-Horizon Optimal Control Problem}

At each time step $k$, MPC requires solving the following finite-horizon
optimal control problem parameterized by the current state $\x_k$
and a task reference $r_k$:

\begin{subequations}
\label{eq:mpc}
\begin{align}
  \min_{\{\x_t\}, \{\uv_t\}}\;  &
    V_f(\x_N;\,r_k) + \sum_{t=0}^{N-1} \ell(\x_t, \uv_t;\,r_k) \label{eq:mpc_obj}\\
  \text{s.t.}\;  &
    \x_{t+1} = f(\x_t) + G(\x_t)\,\uv_t, \quad t = 0, \ldots, N-1,
    \label{eq:mpc_dyn}\\
  & \x_0 = \hat{x}, \label{eq:mpc_ic}\\
  & \umin \leq \uv_t \leq \umax,\quad t = 0, \ldots, N-1, \label{eq:mpc_u}\\
  & \xmin \leq \x_t \leq \xmax,\quad t = 1, \ldots, N, \label{eq:mpc_x}\\
  & A_c(\x_t)\,\uv_t \leq b_c(\x_t),\quad t = 0, \ldots, N-1. \label{eq:mpc_ac}
\end{align}
\end{subequations}

where, with a slight abuse of notation, $t$ denotes the prediction time $k+t$, $\ell : \R^{n_x} \times \R^{n_u} \to \R_{\geq 0}$ is a stage cost and $V_f : \R^{n_x} \to \R_{\geq 0}$ is a terminal cost, both depending on
the task reference $r_k$. 
MPC applies $\uv_k = \uv_0^\star$, i.e., only the first element of the optimal input sequence,
and solve again problem~\eqref{eq:mpc} at the next step $k+1$.

By collecting all the optimization variables into a single vector
\begin{equation}
  \wv = \bigl[\x_0\transpose\ \uv_0\transpose\ \x_1\transpose\ \uv_1\transpose\ 
\ldots\ \x_{N-1}\transpose\ \uv_{N-1}\transpose\ \x_N\transpose\bigr]\transpose
\end{equation}
with  $\wv \in \R^{n_{\wv}}$ and $n_{\wv} = (N+1)n_x + N n_u$,
Problem~\eqref{eq:mpc} can be written more compactly as the NLP
\begin{equation}
  \begin{aligned}
  \wv^\star = \arg\min_{\wv} \; & J(\wv;\,r_k) \\
  \text{s.t.}\quad
    & g_{\mathrm{eq}}(\wv;\,\x_k) = 0, \\
    & \underline{h}\leq h(\wv;\,\x_k) \leq \overline{h}, \\
    & \underline{\wv} \leq \wv \leq \overline{\wv},
  \end{aligned}
  \label{eq:nlp}
\end{equation}
where $J$ is the total cost, $g_{\mathrm{eq}}$ collects all equality constraints~\eqref{eq:mpc_dyn}--\eqref{eq:mpc_ic}, $h$ collects
general inequality constraints affine in the input, 
and $\underline{\wv},\overline{\wv}$ encode the box constraints~\eqref{eq:box_constraints}.

\subsection{KKT Conditions and Optimality Certificates}

The standard first-order KKT conditions for~\eqref{eq:nlp} require stationarity,
primal/dual feasibility, and complementarity. Denoting the Lagrange multipliers
for equality constraints, ranged inequalities, and variable bounds by
$\mum$, $\lam$, and $z_L,z_U\geq 0$ \footnote{Bound
multipliers are split into separate lower and upper components following
the IPOPT convention~\cite{WB06}, which treats them as strictly positive
quantities in the interior-point framework.} respectively, stationarity reads
\begin{equation}
  \nabla_{\wv} \calL(\wv^\star, \mum^\star, \lam^\star, z_L^\star, z_U^\star) = 0,
  \label{eq:kkt_stat}
\end{equation}
where $\calL = J + \mum\transpose g_{\mathrm{eq}} + \lam\transpose h
+ z_L\transpose(\wv_L-\wv) + z_U\transpose(\wv-\wv_U)$.
Primal/dual feasibility and complementarity conditions can be written in smooth form via the
\emph{Fischer--Burmeister} (FB) function~\cite{fischer1992special,lüken2025selfsupervisedlearn}:
\begin{equation}
  \phi_{\mathrm{FB}}(a, b) = \sqrt{a^2 + b^2} - (a + b),
  \label{eq:fb}
\end{equation}
satisfying $\phi_{\mathrm{FB}}(a,b) = 0 \iff a \geq 0,\, b \geq 0,\, ab = 0$.

\subsection{Input-Affine Structure and Single-Step QP}
\label{sec:problem:qp}

The input-affine structure of~\eqref{eq:dynamics} implies that the one-step
state map is \emph{affine} in $\uv_0$ for any fixed $\x_0$:
\begin{equation}
  \x_1 \;=\; f_0 + G_0\,\uv_0,
  \label{eq:x1_affine}
\end{equation}
where $f_0 \in \R^{n_x}$ is the drift under zero input and
$G_0 \in \R^{n_x \times n_u}$ is the control-gain matrix,
both evaluated at $\x_0$.

Every constraint that is affine in $\uv_0$ for a fixed $\x_0$ takes the
form~\eqref{eq:affine_input_constr}.  We distinguish two roles.
\emph{Hard} constraints for input bounds $\umin \leq \uv_0 \leq \umax$ and any
additional affine-in-input bound are encoded directly as
\begin{equation}
  A_c(\x_0)\,\uv_0 \leq b_c(\x_0).
  \label{eq:ineq}
\end{equation}
Instead, we model the bounds $\xmin \leq \x_1 \leq \xmax$ on the first predicted state
as \emph{soft} constraints
\begin{equation}
  A_s(\x_0)\,\uv_0 -s\leq b_s(\x_0), \quad
  A_s = \begin{bmatrix} G_0 \\ -G_0 \end{bmatrix}, \quad
  b_s = \begin{bmatrix} \xmax - f_0 \\ f_0 - \xmin \end{bmatrix}.
  \label{eq:state_ineq}
\end{equation}
where $\sv \in \R^{n_s}_{\geq 0}$ is a vector of non-negative slack variables 
(with $n_s = 2n_x$), which are used to preserve the feasibility of the resulting
\emph{state-dependent QP} in $(\uv_0, \sv)$:
\begin{equation}
  \begin{aligned}
    (\uv_0^\star, \sv^\star) = \arg\min_{\uv_0,\, \sv \geq 0}\; &
      \tfrac{1}{2} \uv_0\transpose H(\x_0)\,\uv_0
      + c(\x_0,r)\transpose\uv_0 + \rho\,1\transpose\sv \\
    \text{s.t.}\quad
    & A_c(\x_0)\,\uv_0 \leq b_c(\x_0), \\
    & A_s(\x_0)\,\uv_0 - \sv \leq b_s(\x_0),
  \end{aligned}
  \label{eq:single_step_qp}
\end{equation}
where $H(\x_0) \succ 0$ and $c(\x_0,r)$ are the parameters of
the quadratic objective, which depend on the state and
reference $r=r_k$, and $\rho > 0$ is the weight associated
with the linear penalty on soft constraint violation.  The joint decision vector
$\zv = [\uv_0\transpose,\sv\transpose]\transpose$ has dimension
$n_z = n_u + n_s$, far smaller than $n_{\wv} = (N+1)n_{x} + N n_{u}$
for typical horizons $N$.

We consider a general stage cost of the form
\begin{equation}
  \ell(\x,\uv;\,r) \;=\;
  \norm{\varphi(\x) - \yref}_{Q}^2 + \norm{\uv - \uref}_{R}^2,
  \label{eq:stage_cost_general}
\end{equation}
where $\varphi:\R^{n_x}\!\to\R^{n_\varphi}$ is a (possibly nonlinear)
output map, $Q \succ 0$, $R \succ 0$ are weight matrices, and $r=[\yref\transpose \uref\transpose]\transpose$ is the task reference.
By linearizing $\varphi(\x_1)$ around $\x_0$ we get the approximation
\begin{equation}
  \varphi(\x_1) \;\approx\;
  \underbrace{\varphi(\x_0) + J_\varphi(\x_0)\,(f_0 - \x_0)}_{\varphi_{\mathrm{drift}}}
  \;+\;
  \underbrace{J_\varphi(\x_0)\,G_0}_{J_\varphi^u}\,\uv_0,
  \label{eq:phi_lin}
\end{equation}
which is linear in $\uv_0$ for any fixed $\x_0$,
where $J_\varphi(\x_0)\in\R^{n_\varphi\times n_x}$ is the Jacobian of
$\varphi$ at $\x_0$.  Substituting~\eqref{eq:phi_lin}
into~\eqref{eq:stage_cost_general} and expanding yields the analytic
baseline parameters:
\begin{subequations}
\begin{equation}
  \begin{aligned}
    H_{\mathrm{an}} &= 2\bigl({J_\varphi^u}^\top Q\,J_\varphi^u + R\bigr), \\
    c_{\mathrm{an}} &= 2{J_\varphi^u}^\top Q\,(\varphi_{\mathrm{drift}} - \yref) - 2R\,\uref.
  \end{aligned}
  \label{eq:qp_params_general}
\end{equation}
When $\varphi(x)=x$, $J_\varphi = I$,
$J_\varphi^u = G_0$, and $\varphi_{\mathrm{drift}} = f_0$,
so~\eqref{eq:qp_params_general} reduces to the familiar quadratic case:
\begin{equation}
  H_{\mathrm{an}} = 2(R + G_0\transpose Q G_0), \quad
  c_{\mathrm{an}} = 2G_0\transpose Q(f_0 - \xref) - 2R\uref.
  \label{eq:qp_params_quad}
\end{equation}
\label{eq:qp_params}
\end{subequations}
In the sequel, we denote by $L_{\mathrm{an}} = \mathrm{chol}(H_{\mathrm{an}})$.

As $(H(\x_0), c(\x_0,r))$ must encode the \emph{entire} $N$-step
value function, not merely the one-step stage cost, while the analytic
baseline~\eqref{eq:qp_params} captures only the linearized
one-step contribution, we need to learn the residual $(\Delta L, \Delta c)$ to account for
the remaining $N{-}1$ horizon steps and, when $\varphi$ is nonlinear,
also for the linearization error. We describe next how to learn them via a neural network.



\section{Amortized QP for Input-Affine NMPC}
\label{sec:method}

\begin{figure*}[ht]
  \centering
  \includegraphics[width=\textwidth]{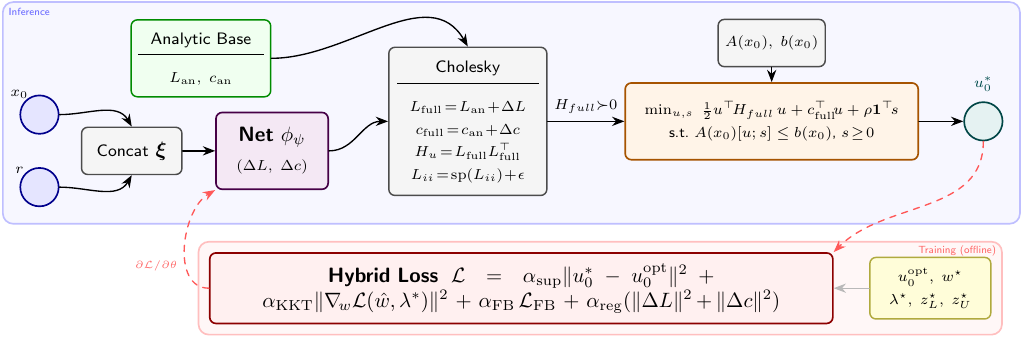}
  \caption{Residual-corrector architecture for amortized input-affine MPC. 
   An analytic baseline $(L_{\mathrm{an}}, c_{\mathrm{an}})$ is computed from the current state and reference; the network $\phi_\psi$ predicts residual corrections $(\Delta L, \Delta c)$.  The Cholesky factor $L_{\mathrm{full}} = L_{\mathrm{an}} + \Delta L$ yields a guaranteed positive-definite Hessian $H_u \succ 0$ after applying softplus to the diagonal.  The resulting QP is solved by a differentiable interior-point layer, producing $\uv_0^*$ with constraint satisfaction.  The network is trained offline on NLP solutions via a hybrid loss combining supervised imitation and KKT-residual penalties.
   }
  \label{fig:arch}
\end{figure*}

As shown in Fig.~\ref{fig:arch}, to learn the residual corrections $(\Delta L, \Delta c)$, we propose the use of the neural network $\phi_\psi$. The network takes the concatenated input
$\xi = [x_0^\top, r^\top]^\top$ and returns $(\Delta L, \Delta c)$ to complement the 
analytic baseline $(L_{\mathrm{an}}, c_{\mathrm{an}})$. 
The corrected Cholesky factor $L_{\mathrm{full}} = L_{\mathrm{an}} + \Delta L$
defines a positive-definite cost matrix $H_u = L_{\mathrm{full}}L_{\mathrm{full}}^\top$,
which together with $c_{\mathrm{full}} = c_{\mathrm{an}} + \Delta c$ and the
state-dependent constraints $A(x_0), b(x_0)$ fully specifies a small QP, to
be solved online to generate the command input $u_0^*$.
The network is trained offline via a hybrid loss combining
supervised imitation of NLP solutions, KKT stationarity, and
Fischer-Burmeister complementarity residuals.

\subsection{QP-Parameter Network}
\label{sec:method:qp_net}

The network $\phi_\psi$ outputs $n_u(n_u+1)/2$ lower-triangular Cholesky
corrections $\Delta L$ and $n_u$ linear cost corrections $\Delta c$,
totalling $n_u(n_u+3)/2$ scalar outputs. The full linear cost is
$c_{\mathrm{full}} = c_{\mathrm{an}} + \Delta c$, and the off-diagonal
Cholesky entries are corrected additively as
$[L_{\mathrm{full}}]_{ij} = [L_{\mathrm{an}}]_{ij} + [\Delta L]_{ij}$
for $i > j$. The diagonal is replaced by
$[L_{\mathrm{full}}]_{ii} = \mathrm{softplus}([L_{\mathrm{an}}]_{ii} +
[\Delta L]_{ii}) + \varepsilon$,
where $\mathrm{softplus}(x) = \log(1+e^x)$ and $\varepsilon > 0$,
guaranteeing strict positivity of the diagonal and hence
$H_{\mathrm{full}} = L_{\mathrm{full}}L_{\mathrm{full}}\transpose \succ 0$.
The constraint matrices $A_c(\x_0)$, $b_c(\x_0)$, $A_s(\x_0)$, $b_s(\x_0)$
are computed analytically from the known dynamics and input-affine structure.
During training, the QP is solved by a batched differentiable
interior-point solver~\cite{tracy2024qpax}, producing $\uv_0$
satisfying~\eqref{eq:ineq} and slack $\sv$ as in~\eqref{eq:state_ineq}.

\subsection{Hybrid Loss Function}
\label{sec:method:loss}

Let $\calD = \{(\x_0^{(i)}, r^{(i)},
\wv^{\star(i)}, \mum^{\star(i)}, \lam^{\star(i)}, z_L^{\star(i)}, z_U^{\star(i)})\}_{i=1}^M$
be an offline dataset of $M$ optimal NLP solutions and their dual variables.
Given the network's predicted first-step control $\hat{\uv}_0^{(i)}$,
we define the \emph{mixed primal vector} $\hat{\wv}^{(i)}$ as the full
optimal solution $\wv^{\star(i)}$ with the first control block and the
resulting one-step-ahead state replaced by the network's prediction:
\begin{equation}
  \hat{\wv}^{(i)} = \wv^{\star(i)}\bigl|_{\substack{
    \uv_0 \leftarrow \hat{\uv}_0^{(i)},\\
    \x_1 \leftarrow f_0^{(i)} + G_0^{(i)}\hat{\uv}_0^{(i)}}}.
    \label{eq:mix_w}
\end{equation}

The network is trained by minimising the following hybrid loss:
$  \calL_{\mathrm{total}} =
    \alpha_{\mathrm{sup}} \calL_{\mathrm{sup}} +
    \alpha_{\mathrm{KKT}} \calL_{\mathrm{KKT}} +
    \alpha_{\mathrm{FB}} \calL_{\mathrm{FB}} +
    \alpha_{\mathrm{reg}} \calL_{\mathrm{reg}},$
where $\alpha_{\mathrm{sup}}, \alpha_{\mathrm{KKT}}, \alpha_{\mathrm{FB}},
\alpha_{\mathrm{reg}} > 0$ are tunable weights, and the individual loss terms are defined as follows.

\paragraph{\normalsize Supervised imitation loss}
\begin{equation}
  \calL_{\mathrm{sup}} =
    \frac{1}{M}\sum_{i=1}^M \norm{\hat{\uv}_0^{(i)} - \uv_0^{\star(i)}}^2
    \label{eq:loss_sup}
\end{equation}
penalizing the deviation from the optimal first control action. 

\paragraph{\normalsize KKT stationarity loss}
To enforce~\eqref{eq:kkt_stat}, we penalize
\begin{equation}
  \calL_{\mathrm{KKT}} =
    \frac{1}{M}\sum_{i=1}^M
      \norm{\nabla_{\wv} \calL(\hat{\wv}^{(i)},\, \mum^{\star(i)},\, \lam^{\star(i)},\,
             z_L^{\star(i)},\, z_U^{\star(i)})}^2
    \label{eq:loss_kkt}
\end{equation}
at the optimal dual variables from the dataset, measuring the stationarity residual of the network's prediction
$\hat{\wv}^{(i)}$ for the full NLP.

\paragraph{\normalsize Fischer–Burmeister complementarity loss}
The violation of the complementarity conditions and inequality constraints 
are penalized via the FB function~\eqref{eq:fb}. For box constraints on $\wv$, we set
$\calL_{\mathrm{FB}}^{\mathrm{box}} = \frac{1}{M}\sum_{i=1}^M \Bigl(
      \norm{\phi_{\mathrm{FB}}(\hat{\wv}^{(i)} - \underline{\wv},\;z_L^{\star(i)})}^2 
      + \norm{\phi_{\mathrm{FB}}(\overline{\wv} - \hat{\wv}^{(i)},\;z_U^{\star(i)})}^2
      \Bigr)$.

For ranged inequality constraints $\underline{h_j} \leq h_j(\wv) \leq \overline{h_j}$
with signed multiplier $\lambda_j^\star$ (positive when the upper bound is
active, negative when the lower bound is active), we penalize instead
$\calL_{\mathrm{FB}}^{\mathrm{ineq}} = \frac{1}{M}\sum_{i=1}^M \sum_j \Bigl(
      \norm{\phi_{\mathrm{FB}}(\overline{h_j} - h_j(\hat{\wv}^{(i)}),\,
            [\lambda_j^{\star(i)}]^+)}^2
      + \norm{\phi_{\mathrm{FB}}(h_j(\hat{\wv}^{(i)}) - \underline{h_j},\,
            [\lambda_j^{\star(i)}]^-)}^2
    \Bigr)$.

The total FB loss is
$\calL_{\mathrm{FB}} = \calL_{\mathrm{FB}}^{\mathrm{box}} +
\calL_{\mathrm{FB}}^{\mathrm{ineq}}$.
The optimal multipliers in $\calD$ are treated again as fixed constants during
training.

\paragraph{\normalsize Regularization towards the analytic baseline}
To prevent the learned corrections from deviating too far from the analytic baseline and regularize the learning problem, we include a small $\ell_2$ penalty on the residuals:
$  \calL_{\mathrm{reg}} = \frac{1}{M}\sum_{i=1}^M
    \bigl(\norm{\Delta L^{(i)}}^2 + \norm{\Delta c^{(i)}}^2\bigr)$.
The weight $\alpha_{\mathrm{reg}}$ on $\calL_{\mathrm{reg}}$ is set to a small value. 

Note that $\hat{\wv}^{(i)}$ in~\ref{eq:mix_w} satisfies the dynamic equations only at the first step when
transitioning from $x_0$ to $x_1$; the dynamics at $t\geq 1$ are violated when $\hat{\uv}_0 \!\ne\! \uv_0^\star$,
as the subsequent states $\x_2, \ldots, \x_N$ are still taken from the optimal solution $\wv^\star$ and thus are consistent with $\uv_0^\star$ rather than $\hat{\uv}_0$. Therefore, the KKT and FB losses are evaluated at a dynamically infeasible point, accurate to first order in $\hat{\uv}_0 - \uv_0^\star$.

\section{Case Study: Planar Robotic Arm}
\label{sec:experiments}

\begin{figure*}[htbp]
  \centering
  \includegraphics[width=\textwidth]{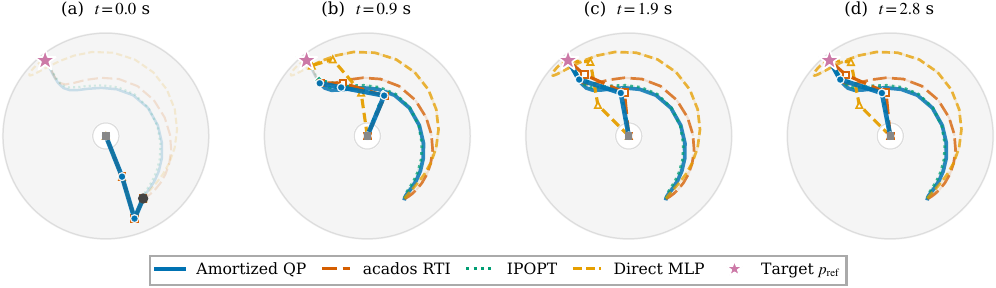}
  \caption{End-effector trajectories in the Cartesian workspace for the
           amortized QP (solid blue), RTI (dashed orange), IPOPT (dashed
           green), and direct MLP (dashed yellow), from a common initial
           configuration (filled circle) to the target $p_{\rm ref}$
           (star). All solvers successfully reach the target; only the amortized QP closely matches the IPOPT trajectory.}
  \label{fig:workspace}
\end{figure*}

\subsection{System Model}

We test the proposed nonlinear MPC approximation method on a three-link planar robotic arm operating in the vertical plane under gravity, with link lengths
$l = [1.0,\,1.0,\,0.5]$\,m, masses $m = [1.0,\,1.0,\,0.5]$\,kg,
centre-of-mass offsets $l_c = [0.5,\,0.5,\,0.25]$\,m, and moments of
inertia $I = [0.0833,\,0.0833,\,0.0104]$\,kg\,m$^2$.
The system is described by the Euler--Lagrange equations
\begin{equation}
  M(q)\,\ddot{q} + C(q,\dot{q})\,\dot{q} + g(q) = \tau,
  \label{eq:euler_lagrange}
\end{equation}
where $M(q)\in\R^{3\times3}$ is the inertia matrix, $C(q,\dot{q})$ the
Coriolis matrix, $g(q)$ the gravity vector, and $\tau\in\R^3$ the joint
torques.

Choosing joint accelerations $\uv=\ddot{q}\in\R^3$ as the control input yields
an input-affine system~\eqref{eq:dynamics} with state
$\x=[q\transpose,\dot{q}\transpose]\transpose\in\R^6$.
Both the MPC prediction model and the plant model are integrated with a
fixed-step RK4 scheme; the plant model is simulated using the full nonlinear
dynamics~\eqref{eq:euler_lagrange}, while the prediction model uses the simplified
double-integrator model $\ddot{q}=\uv$. Torque signals are recovered via
inverse dynamics $\tau = M(q)\uv + C(q,\dot{q})\dot{q} + g(q)$.

The affine one-step map~\eqref{eq:x1_affine} used to build the QP
constraints is obtained by applying zero-order hold to the double integrator,
giving the exact discrete matrices
\begin{equation}
  G_0 = \begin{bmatrix}\tfrac{1}{2}T_s^2 I_{n_q}\\T_s I_{n_q}\end{bmatrix},
  \qquad
  f_0 = \begin{bmatrix}q + T_s\dot{q}\\\dot{q}\end{bmatrix}.
  \label{eq:arm_zoh}
\end{equation}

\subsection{MPC Problem Setup}

We setup the MPC problem~\eqref{eq:mpc} with horizon $N = 15$ and the
\emph{Cartesian end-effector tracking} objective
\begin{equation}
  \ell(\x, \uv;\,p_{\mathrm{ref}}) =
    \norm{p(q) - p_{\mathrm{ref}}}_{W_{ee}}^2 +
    \norm{\dot{q}}_{W_{dq}}^2 +
    \norm{\uv}_{W_u}^2,
  \label{eq:cart_stage_cost}
\end{equation}
where $p(q) \in \R^2$ is the Cartesian end-effector position (forward
kinematics), $p_{\mathrm{ref}} \in \R^2$ is the Cartesian target, and
$W_{ee} = \mathrm{diag}(150, 150)$, $W_{dq} = 0.5\,I_3$,
$W_u = 0.02\,I_3$.  The terminal cost uses $W_{ee,f} = 20\,W_{ee}$,
$W_{dq,f} = 10\,W_{dq}$.  Here $r = p_{\mathrm{ref}}$.

The following constraints are enforced at every horizon step:
\emph{box constraints} on joint positions $|q_i| \leq \pi$\,rad,
joint velocities $|\dot{q}_i| \leq 5$\,rad/s, and
accelerations $|u_i| \leq 20$\,rad/s$^2$ for all $i=1,2,3$;
\emph{affine-in-input constraints}~\eqref{eq:affine_input_constr}, with joint
torques expressed via inverse dynamics~\eqref{eq:euler_lagrange},
$|\tau_i| \leq 50$\,N\,m for joints 1--2, and $|\tau_3| \leq 20$\,N\,m
for the distal link.


The stage cost~\eqref{eq:cart_stage_cost} is \emph{nonlinear} in $\x$ (through
the forward kinematics $p(q)$), so the closed-form
parameters~\eqref{eq:qp_params_quad} do not apply directly.  We derive an
analytic baseline by linearising the end-effector position around the current
configuration:
\begin{equation*}
  p(q_1) \approx \underbrace{p(q) + J_{ee}(q)\,T_sI_3\dot{q}}_{%
    p_{\mathrm{drift}}}
  + \underbrace{J_{ee}(q)\,\tfrac{1}{2}T_s^2 I_3}_{J_{ee}^u}\,\uv,
  \label{eq:ee_lin}
\end{equation*}
where $J_{ee}(q) \in \R^{2\times 3}$ is the end-effector Jacobian and
$p_{\mathrm{drift}}$ is the drift under zero control.  Substituting
into~\eqref{eq:cart_stage_cost} and expanding yields
\begin{align}
  H_{\mathrm{an}} &= 2\bigl(J_{ee}^{u\top} W_{ee} J_{ee}^u
                          + T_s^2 W_{dq}
                          + W_u\bigr), \label{eq:Han}\\
  c_{\mathrm{an}} &= 2 J_{ee}^{u\top} W_{ee}(p_{\mathrm{drift}} - p_{\mathrm{ref}})
                          + 2T_s W_{dq}\dot{q}. \label{eq:can}
\end{align}


The hard constraints~\eqref{eq:ineq} encode the torque bounds
$\tau = M(q)\uv + h(q,\dot{q}) \in [\underline{\tau},
\overline{\tau}]$ together with the acceleration bounds:
\begin{equation}
  A_c(\x) = \begin{bmatrix} M(q) \\ -M(q) \\ I_3 \\ -I_3 \end{bmatrix}\!,
  \quad
  b_c(\x) = \begin{bmatrix}
    \overline{\tau} - h \\
    h - \underline{\tau} \\
    \umax \\ -\umin
  \end{bmatrix}\!.
  \label{eq:arm_hard}
\end{equation}
The one-step-ahead state is exact under the ZOH model~\eqref{eq:arm_zoh}:
$q_1 = q + T_s\dot{q} + \tfrac{1}{2}T_s^2\uv$,
$\dot{q}_1 = \dot{q} + T_s\uv$.  Stacking the four upper/lower
bound pairs gives the arm-specific instance of~\eqref{eq:state_ineq} using
$G_0$ from~\eqref{eq:arm_zoh}:
\begin{equation}
  A_s =
  \begin{bmatrix}
    \tfrac{1}{2}T_s^2 I_3 \\ -\tfrac{1}{2}T_s^2 I_3 \\
    T_s I_3 \\ -T_s I_3
  \end{bmatrix}\!,
  \;
  b_s(\x) =
  \begin{bmatrix}
    \overline{q} - (q + T_s\dot{q}) \\
    (q + T_s\dot{q}) - \underline{q} \\
    \overline{\dot{q}} - \dot{q} \\
    \dot{q} - \underline{\dot{q}}
  \end{bmatrix}\!,
  \label{eq:arm_soft}
\end{equation}
yielding $n_s = 4 n_u = 12$ slack variables.  The QP~\eqref{eq:single_step_qp}
therefore has $n_z = n_u + n_s = 15$ decision variables and $36$ inequality
rows ($12$ hard from~\eqref{eq:arm_hard}, $12$ soft from~\eqref{eq:arm_soft},
and $12$ non-negativity rows $-\sv \leq 0$).

\subsection{Data Generation and Training}

Trajectories are generated under warm-started IPOPT~\cite{WB06} with
$\epsilon_{\mathrm{tol}}=10^{-6}$ on all termination criteria, $T_s=0.05$\,s,
and $T_{\mathrm{sim}}=0.005$\,s. Each trajectory runs up to $80$ steps from
an independently drawn $(\x_0,p_{\mathrm{ref}})$; those failing Cartesian
($0.04$\,m) or velocity ($0.08$\,rad/s) tolerances are discarded.
$(q,\dot{q},p_{\mathrm{ref}})\in\R^8$ are drawn via Latin hypercube sampling:
$q\sim\mathcal{U}([-\pi,\pi]^3)$, $\dot{q}\sim\mathcal{U}([-5,5]^3)$,
$p_{\mathrm{ref}}$ over the reachable annulus $\rho\in[0.30,2.35]$\,m,
yielding $M=100\,000$ samples.

$\phi_\psi$ is a three-hidden-layer MLP ($[128,128,64]$, ReLU) mapping
$\xi\in\R^8$ to the $9$-dimensional residual $(\Delta L,\Delta c)$ over
the analytic baseline~\eqref{eq:Han}--\eqref{eq:can}, initialized to zero
so the QP recovers the analytic solution at startup. Training uses
Adam~\cite{KB14} with cosine decay ($10^{-3}\!\to\!10^{-5}$), $200$ epochs,
batch size $256$, $(\alpha_{\mathrm{sup}},\alpha_{\mathrm{KKT}},
\alpha_{\mathrm{FB}},\alpha_{\mathrm{reg}})=(1,10^{-4},10^{-3},10^{-4})$,
$\rho=1000$, $\varepsilon=10^{-3}$, with QP gradients via implicit
differentiation in JAX~\cite{jax2018github}.

\begin{remark}
$\alpha_{\mathrm{KKT}}$ and $\alpha_{\mathrm{FB}}$ are kept small
relative to $\alpha_{\mathrm{sup}}$: both losses are evaluated at the
mixed primal $\hat{\wv}^{(i)}$, giving a gradient approximation
unreliable far from the NLP solution, and the KKT
gradient~\eqref{eq:loss_kkt} is numerically larger in scale than the
supervised loss. The KKT and FB terms thus serve as a complementary
signal near constraint boundaries once the supervised term has converged.
\end{remark}

\subsection{Numerical Setup}
All evaluations are run on an Intel Core i5-9400H mobile CPU.
We evaluate the trained policy in closed-loop against three baselines: an
\emph{acados} SQP-RTI solver~\cite{acados} performing one Gauss--Newton
iteration per step; full-horizon IPOPT as an optimality reference; and a
\emph{direct MLP} that shares the same architecture and composite loss as
the amortized policy but omits the differentiable QP layer, directly
predicting $\uv_0^\star$ from $\xi$. Each scenario draws $q_0$ uniformly
in $[-\pi,\pi]^3$ at rest ($\dot{q}_0=0$) and a target $p_{\mathrm{ref}}$
from the reachable annulus $r\in[0.30,2.35]$\,m; all controllers drive
the nonlinear plant for $4$\,s (Figure~\ref{fig:workspace}). Because
IPOPT solves a nonconvex NLP it may fail to converge to a global minimum;
such scenarios are discarded and replaced so that IPOPT can be used to get
a lower bound on the closed-loop MPC performance: $ J_{\mathrm{cl}} = \sum_{k=0}^{T-1} \ell(x_k, u_k) + \ell_T(x_T).$
We collect $100$ valid scenarios under this protocol.

\subsection{Results}

\begin{figure}
  \centering
  \includegraphics[width=\columnwidth]{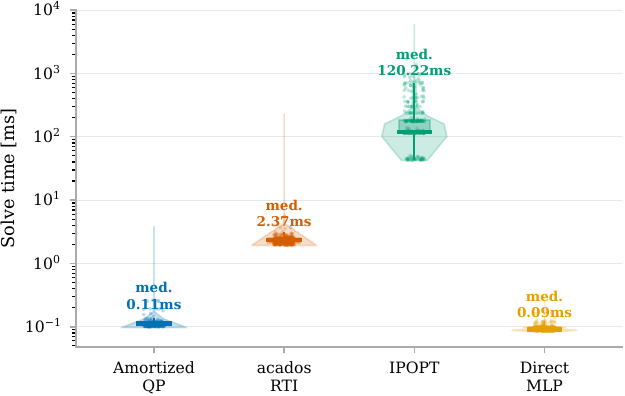}
  \caption{Per-step solve time. The amortized policy (blue) is
           ${\approx}21\times$ faster than RTI (orange) and
           ${\approx}1826\times$ faster than IPOPT (purple). The direct
           MLP (red) is slightly faster, but this comes at the expense of
           constraint satisfaction and convergence reliability. Median solve
           times are reported above each distribution.}
  \label{fig:solve_time}
\end{figure}

Tables~\ref{tab:results_perf}--\ref{tab:results_cost} and
Figure~\ref{fig:solve_time} show the obtained results in terms of closed-loop performance and solution time across scenarios.

The amortized policy converges on all $100$ scenarios and achieves a mean
closed-loop cost within $4.7\%$ of the IPOPT optimum, while acados RTI
diverges on $12$ scenarios and sits $17.5\%$ above on average (up to
$6.9\times$ in the worst case). The direct MLP converges on only $73/100$
scenarios and sits $12.0\%$ above IPOPT, with hard constraint violations
reaching $46.3$\,N\,m in torque and $20.8$\,rad/s$^2$ in acceleration;
this confirms that the QP layer is essential for both feasibility and
reliability. The amortized policy runs in $0.116$\,ms mean, a $21\times$ speedup over RTI and over $1000\times$ over IPOPT.
 Hard bounds on torque and acceleration are satisfied to numerical precision. Soft joint-velocity
constraints are mildly violated during transients, with smaller peak
excursions for the amortized policy ($1.42$ vs.\
$3.30$\,rad/s for RTI). Slack is active in only $0.7\%$ of steps,
confirming that constraint softening plays a minor role in nominal
operation.

\begin{table}
  \centering
  \caption{Computational efficiency and tracking performance ($100$
           scenarios, $T=80$ steps)}
  \label{tab:results_perf}
  \renewcommand{\arraystretch}{1}
  \setlength{\tabcolsep}{3pt}
  \small
  \begin{tabular}{lcccc}
    \toprule
    Metric & Amort. & RTI & IPOPT & MLP \\
    \midrule
    Mean time (ms)         & $0.116$   & $2.453$   & $211.8$      & $0.095$   \\
    Max time (ms)          & $3.780$   & $224.4$   & $5837$       & $0.284$   \\
    Speedup vs.\ IPOPT$^{*}$    & $1826$    & $86$      & ---          & $2229$    \\
    Speedup vs.\ RTI$^{*}$      & $21$      & ---       & ---          & $26$      \\
    \midrule
    Convergence            & $100/100$ & $88/100$  & $100/100^{\dagger}$ & $73/100$ \\
    Mean EE err.\ (m)      & $1.5\text{e}{-3}$ & $4.2\text{e}{-2}$ & $1.2\text{e}{-3}$ & $4.0\text{e}{-2}$ \\
    Max EE err.\ (m)       & $2.8\text{e}{-2}$ & $1.163$           & $3.4\text{e}{-2}$ & $6.6\text{e}{-1}$ \\
    \bottomrule
    \multicolumn{5}{l}{${}^{*}$Ratio of mean solve times.} \\
    \multicolumn{5}{l}{${}^{\dagger}$Non-converging scenarios discarded by construction.}
  \end{tabular}
\end{table}

\begin{table}
  \centering
  \caption{Closed-loop cost and constraint satisfaction ($100$ scenarios).
           $J_{\mathrm{cl}}$ ratios relative to IPOPT.}
  \label{tab:results_cost}
  \renewcommand{\arraystretch}{1}
  \setlength{\tabcolsep}{4pt}
  \small
  \begin{tabular}{lcccc}
    \toprule
    Metric & Amort. & RTI & IPOPT & MLP \\
    \midrule
    Mean $J_{\mathrm{cl}}$ & $3642.5$  & $4273.0$  & $3574.9$  & $3878.2$  \\
    Max $J_{\mathrm{cl}}$  & $23451$   & $29620$   & $23239$   & $26496$   \\
    Mean ratio             & $1.047$   & $1.175$   & $1.000$   & $1.120$   \\
    Max ratio              & $1.596$   & $6.914$   & $1.000$   & $2.633$   \\
    \midrule
    Max $\tau$ viol.\ (N\,m)     & $0$    & $10^{-11}$ & $10^{-7}$ & $46.3$ \\
    Max $u$ viol.\ (rad/s$^2$)   & $0$    & $10^{-15}$ & $10^{-7}$ & $20.8$ \\
    Max $\dot{q}$ viol.\ (rad/s) & $1.42$ & $3.30$     & $2.65$    & $2.56$ \\
    Slack active (\%)            & $0.7$  & ---        & $4.3$     & ---    \\
    \bottomrule
  \end{tabular}
\end{table}

\section{Conclusion}
\label{sec:conclusion}
We have presented a learning-based amortized framework for MPC of input-affine nonlinear
systems that only requires solving a state-dependent QP online. We showed that, by 
using a hybrid loss combining supervised imitation with physics-informed
KKT-residual penalties, we get superior performance and feasibility compared to merely fitting collected nonlinear MPC data.  Moreover, our approach can achieve considerable speedups over RTI and orders-of-magnitude speedups over full-horizon NLP solvers while maintaining comparable tracking performance.


\bibliographystyle{IEEEtran}
\bibliography{Ref}

\begin{thebibliography}{10}
\providecommand{\url}[1]{#1}
\csname url@samestyle\endcsname
\providecommand{\newblock}{\relax}
\providecommand{\bibinfo}[2]{#2}
\providecommand{\BIBentrySTDinterwordspacing}{\spaceskip=0pt\relax}
\providecommand{\BIBentryALTinterwordstretchfactor}{4}
\providecommand{\BIBentryALTinterwordspacing}{\spaceskip=\fontdimen2\font plus
\BIBentryALTinterwordstretchfactor\fontdimen3\font minus \fontdimen4\font\relax}
\providecommand{\BIBforeignlanguage}[2]{{%
\expandafter\ifx\csname l@#1\endcsname\relax
\typeout{** WARNING: IEEEtran.bst: No hyphenation pattern has been}%
\typeout{** loaded for the language `#1'. Using the pattern for}%
\typeout{** the default language instead.}%
\else
\language=\csname l@#1\endcsname
\fi
#2}}
\providecommand{\BIBdecl}{\relax}
\BIBdecl

\bibitem{rawlings2017model}
J.~B. Rawlings, D.~Q. Mayne, and M.~Diehl, \emph{Model Predictive Control: Theory, Computation, and Design}, 2nd~ed.\hskip 1em plus 0.5em minus 0.4em\relax Nob Hill Publishing, 2017.

\bibitem{ferreau2014qpoases}
H.~J. Ferreau, C.~Kirches, A.~Potschka, H.~G. Bock, and M.~Diehl, ``qpoases: A parametric active-set algorithm for quadratic programming,'' \emph{Mathematical Programming Computation}, vol.~6, no.~4, pp. 327--363, 2014.

\bibitem{ODYSQP}
G.~Cimini, A.~Bemporad, and D.~Bernardini, ``{ODYS QP Solver},'' {ODYS S.r.l. (\nobreak{\url{https://odys.it/qp}})}, Sep. 2017.

\bibitem{frison2020hpipm}
G.~Frison and M.~Diehl, ``Hpipm: a high-performance quadratic programming framework for model predictive control,'' \emph{IFAC-PapersOnLine}, vol.~53, no.~2, pp. 6563--6569, 2020.

\bibitem{bemporad2002explicit}
A.~Bemporad, M.~Morari, V.~Dua, and E.~N. Pistikopoulos, ``The explicit linear quadratic regulator for constrained systems,'' \emph{Automatica}, vol.~38, no.~1, pp. 3--20, 2002.

\bibitem{diehl2002real}
M.~Diehl, H.~G. Bock, and J.~P. Schl{\"o}der, ``A real-time iteration scheme for nonlinear optimization in optimal feedback control,'' \emph{SIAM Journal on Control and Optimization}, vol.~43, no.~5, pp. 1714--1736, 2005.

\bibitem{gros2020linear}
S.~Gros, M.~Zanon, R.~Quirynen, A.~Bemporad, and M.~Diehl, ``From linear to nonlinear {MPC}: Bridging the gap via the real-time iteration,'' \emph{International Journal of Control}, vol.~93, no.~1, pp. 62--80, 2020.

\bibitem{amos2022tutorial}
B.~Amos, ``Tutorial on amortized optimization,'' \emph{Foundations and Trends in Machine Learning}, vol.~16, no.~5, pp. 592--732, 2023.

\bibitem{AZB24}
S.~Abdufattokhov, M.~Zanon, and A.~Bemporad, ``Learning {Lyapunov} terminal costs from data for complexity reduction in nonlinear model predictive control,'' \emph{Int. Journal of Robust and Nonlinear Control}, vol.~34, no.~13, pp. 8676--8691, 2024.

\bibitem{SBB25}
M.~Schaller, A.~Bemporad, and S.~Boyd, ``\href{https://stanford.edu/~boyd/papers/pdf/lpcf.pdf}{Learning Parametric Convex Functions},'' 2025, \url{http://arxiv.org/abs/2506.04183}.

\bibitem{wabersich2021predictive}
K.~P. Wabersich and M.~N. Zeilinger, ``A predictive safety filter for learning-based control of constrained nonlinear dynamical systems,'' \emph{Automatica}, vol. 129, p. 109597, 2021.

\bibitem{sambharya2023end}
R.~Sambharya, G.~Hall, B.~Amos, and B.~Stellato, ``Learning to warm-start fixed-point optimization algorithms,'' \emph{J. Mach. Learn. Res.}, vol.~25, no.~1, Jan. 2024.

\bibitem{amos2017optnet}
B.~Amos and J.~Z. Kolter, ``{OptNet}: Differentiable optimization as a layer in neural networks,'' in \emph{Proc. International Conference on Machine Learning (ICML)}, 2017, pp. 136--145.

\bibitem{WB06}
A.~W{\"a}chter and L.~T. Biegler, ``On the implementation of an interior-point filter line-search algorithm for large-scale nonlinear programming,'' \emph{Mathematical programming}, vol. 106, no.~1, pp. 25--57, 2006.

\bibitem{fischer1992special}
A.~Fischer, ``A special {Newton}-type optimization method,'' \emph{Optimization}, vol.~24, no. 3--4, pp. 269--284, 1992.

\bibitem{lüken2025selfsupervisedlearn}
\BIBentryALTinterwordspacing
L.~Lüken and S.~Lucia, ``Self-supervised learning of iterative solvers for constrained optimization,'' 2025. [Online]. Available: \url{https://arxiv.org/abs/2409.08066}
\BIBentrySTDinterwordspacing

\bibitem{tracy2024qpax}
K.~Tracy and Z.~Manchester, ``On the differentiability of the primal-dual interior-point method,'' 2024.

\bibitem{KB14}
D.~P. Kingma and J.~Ba, ``Adam: A method for stochastic optimization,'' \emph{arXiv preprint 1412.6980}, 2014.

\bibitem{jax2018github}
\BIBentryALTinterwordspacing
J.~Bradbury, R.~Frostig, P.~Hawkins, M.~J. Johnson, C.~Leary, D.~Maclaurin, G.~Necula, A.~Paszke, J.~Vander{P}las, S.~Wanderman-{M}ilne, and Q.~Zhang, ``{JAX}: composable transformations of {P}ython+{N}um{P}y programs,'' 2018. [Online]. Available: \url{http://github.com/jax-ml/jax}
\BIBentrySTDinterwordspacing

\bibitem{acados}
R.~Verschueren, G.~Frison, D.~Kouzoupis, J.~Frey, N.~v. Duijkeren, A.~Zanelli, B.~Novoselnik, T.~Albin, R.~Quirynen, and M.~Diehl, ``acados -- a modular open-source framework for fast embedded optimal control,'' \emph{Mathematical Programming Computation}, vol.~14, no.~1, pp. 147--183, 2022.

\end{thebibliography}

\end{document}